\documentclass[fleqn,usenatbib]{mnras}

\usepackage{newtxtext,newtxmath}

\usepackage[T1]{fontenc}
\usepackage{ae,aecompl}

\usepackage{graphicx}	
\usepackage{color}

\title[GW bursts from single cosmic strings]{Searching for gravitational wave bursts from cosmic string cusps with the Parkes Pulsar Timing Array}

\author[Yonemaru et al.]{N. Yonemaru,$^{1,2}$\thanks{E-mail: naoyuki.yonemaru@gmail.com}
S. Kuroyanagi,$^{3,4}$
G. Hobbs,$^{2}$
K. Takahashi,$^{1,5}$
X.-J. Zhu,$^{6,7}$ \newauthor
W. A. Coles,$^{8}$
S. Dai, $^{2}$
{E. Howard},$^{2,9,10}$
{R. Manchester},$^{2}$
{D. Reardon},$^{7,11}$ 
{C. Russell},$^{12}$\newauthor
R. Shannon,$^{7,11}$
{N. Thyagarajan},$^{13}$
R. Spiewak, $^{7,11}$
{J.-B. Wang}$^{14,15}$
\\
$^{1}$Kumamoto University, Graduate School of Science and Technology, Kumamoto, 860-8555, Japan\\
$^{2}$CSIRO Astronomy and Space Science, PO Box 76, Epping NSW 1710, Australia\\
$^{3}$Nagoya University, Graduate School of Science, Nagoya, 464-8601, Japan\\
$^{4}$Instituto de F\'isica Te\'orica UAM-CSIC, Universidad Auton\'oma de Madrid, Cantoblanco, 28049 Madrid, Spain\\
$^{5}$International Research Organization for Advanced Science and Technology, Kumamoto University, Kumamoto 860-8555, Japan\\
$^{6}$School of Physics and Astronomy, Monash University, VIC 3800, Australia\\
$^{7}$OzGrav: The Australian Research Council Centre of Excellence for Gravitational Wave Discovery\\
$^{8}$Electrical and Computer Engineering, University of California at San Diego, La Jolla, California, USA \\
$^{9}$Department of Physics and Astronomy, Macquarie University, Sydney, 2109, Australia\\
$^{10}$Centre for Quantum Dynamics, Griffith University, Brisbane, Queensland, 4111, Australia \\
$^{11}$Centre for Astrophysics and Supercomputing, Swinburne University of Technology, P.O. Box 218, Hawthorn, VIC 3122, Australia\\
$^{12}$CSIRO Scientific Computing Services, Australian Technology Park, Locked Bag 9013, Alexandria, NSW 1435, Australia \\
$^{13}$National Radio Astronomy Observatory, 1003 Lopezville Rd, Socorro, New Mexico 87801, USA\\
$^{14}$Xinjiang Astronomical Observatory, Chinese Academy of Science, 150 Science 1-Street, Urumqi, Xinjiang, China, 830011 \\
$^{15}$Key Laboratory of Radio Astronomy, Chinese Academy of Science, 150 Science 1-Street, Urumqi, Xinjiang, China, 830011 \\
}

\date{Accepted XXX. Received YYY; in original form ZZZ}

\pubyear{2019}

\begin{document}
\label{firstpage}
\pagerange{\pageref{firstpage}--\pageref{lastpage}}
\maketitle

\begin{abstract}

Cosmic strings are potential gravitational wave (GW) sources that can be probed by pulsar timing arrays (PTAs). In this work we develop a detection algorithm for a GW burst from a cusp on a cosmic string, and apply it to Parkes PTA data. We find four events with a false alarm probability less than 1\%. However further investigation shows that all of these are likely to be spurious. As there are no convincing detections we place upper limits on the GW amplitude for different event durations. From these bounds we place limits on the cosmic string tension of $G\mu \sim 10^{-5}$, and highlight that this bound  is independent from those obtained using other techniques. We discuss the physical implications of our results and the prospect of probing cosmic strings in the era of Square Kilometre Array (SKA). 
\end{abstract}

\begin{keywords}
gravitational waves -- pulsars: general -- methods: data analysis
\end{keywords}



\section{Introduction}
Pulsar timing arrays (PTAs) are a sensitive probe of low-frequency  ($10^{-9}$ to $10^{-7}$\,Hz) gravitational waves (GWs).  The PTAs regularly monitor the times of arrival (ToAs) of pulses from a large number of stable millisecond pulsars \citep{Foster}. GWs affect the pulse propagation and change the pulse ToAs (see, e.g., \citealt{estabrook75}, \citealt{sazhin78} and \citealt{detweiler79}). The deviations of the ToAs from those expected from the pulsar timing model \citep{Edwards} are called ``timing residuals". Timing residuals occur for numerous reasons (including the presence of a GW signal) and numerous publications describe how the signal of a GW could be detected (such as \citealt{spolaor19} and references therein). Currently, several PTAs are in operation including the Parkes PTA in Australia (PPTA; \citealt{Manchester2012}), the European PTA (EPTA; \citealt{Kramer2013}) and NANOGrav in North America \citep{nanograv}. These PTAs share data as part of the International Pulsar Timing Array (IPTA) project (see, e.g., \citealt{perera19}). Groups in China, India and South Africa are now also joining the IPTA efforts using their radio telescopes. Within a decade, the square kilometre array (SKA) will be in operation \citep{Kramer2015}, which will further improve the sensitivity of PTAs. 

The GW frequency range detectable by pulsar timing is determined by the observational time span and cadence.  The main sources of GWs in the nano-hertz regime are inspiraling supermassive black hole (SMBH) binaries (e.g., \citealt{spolaor19}). Their incoherent superposition produces a GW background. The PPTA has placed upper bounds for the stochastic GW background \citep{Shannon}, continuous GWs from single SMBH binaries \citep{Zhu}, the GW memory \citep{Wang} and the density of ultralight scalar-field dark matter in the Galaxy \citep{Porayko}. GWs from cosmic strings are potentially detectable in the PTA frequency band. Constraints on cosmic strings have been discussed using an upper bound on the GW background by \cite{Lentati2015} with the EPTA 18-year data set and by \cite{Arzoumanian} with the NANOGrav 11-year data set. 

Cosmic strings are hypothetical, stable, macroscopic, one-dimensional objects of high energy density formed during a spontaneous symmetry-breaking phase transition in the early Universe. Also known as topological defects, they are predicted by quantum field theory and condensed-matter models with implications for super-symmetric unified field theories, such as D-brane models. They were first proposed as a possible explanation for early-Universe structure formation \citep{Kibble,Vilenkin,Sarangi,Jones,Dvali} with symmetry breaking at the grand unification scale. Several cosmological inflationary models based on superstring theory predicted defects such as fundamental strings, D-strings or bound states at cosmological scale that may have left signatures in the early Universe through their nonlinear evolution. 

Since strings are predicted to emit GW bursts,  experiments such as the PTAs provide a means to test for their existence. The strongest GW bursts are produced at singularities on string loops termed cusps \citep{Damour}. A cusp represents a highly Lorentz-boosted region in the loop that emits a strong beam of GWs, also called a GW burst. Usually numerous bursts overlap one another and form a GW background. However, in a specific region of parameter space, a small number of GW bursts could dominate the background \citep{Kuroyanagi:2016ugi} and, if the GW amplitude of the event is strong enough, it may be observed as a single burst. Individual GWs from cosmic strings have been searched in the direction of fast radio burst sources \citep{Abbott2016} and with the Laser Interferometer Gravitational-wave Observatory (LIGO) \citep{Abbott2009,Abbott2018}, but not previously by the PTAs (although methods to search for burst events have been described by \citealt{finn10} and \citealt{madison16}). In this paper, we perform the first search for an individual GW burst from a cosmic string, using the second data set release from the Parkes PTA (PPTA) project. The waveform of a GW burst from a cusp has a unique shape regardless of the string tension and loop size. We have prepared waveform templates for the GW burst from a cusp and performed a matched filter search with a global fit. 

This paper is organized as follows. In section \ref{sec:Obs}, we briefly introduce the PPTA data set and its noise properties. In section \ref{sec:CS signal}, we describe the expected timing residual caused by the GW burst from a cosmic string cusp. In section \ref{sec:principle}, we present our detection algorithm to search for the GW burst. In section \ref{sec:results}, we apply our algorithm to the PPTA data and present the results. In section \ref{sec:discussion}, we place constraints on the cosmic string tension, and discuss the physical implications of our results.

\section{Observation data set} \label{sec:Obs}

We use the same PPTA data set as \cite{Porayko}. The observing systems and data processing techniques are similar to the first PPTA data release (DR1) as described in \cite{Manchester2012}. In brief, the data set includes 26 millisecond pulsars observed at intervals of 2 to 3 weeks between 2004 and 2016 using the Parkes telescope. The data set is available from the CSIRO pulsar data archive\footnote{data.csiro.au} at \url{https://doi.org/10.25919/5bc67e4b7ddf2}. 

Before the GW search, we fitted the pulsar ToAs with a timing model and formed timing residuals using the standard \textsc{tempo2} software package \citep{Hobbs2006}. Parameters of this timing model include the pulsar sky location, spin frequency and spin-down rate, dispersion measure, proper motion, parallax and (when applicable) binary orbital parameters. Additionally, constant offsets are fitted among ToAs collected with different receiver/signal processor systems. 

The timing residuals include measurement errors in estimating the ToAs, which are independent (white) noise, and also unmodelled low-frequency (red) noise, which is due to irregularities in the pulse emission. Both noise sources affect the estimation of the pulsar parameters and subsequently our search for the cosmic string signals. We have updated the noise modelling of the data set of \cite{Porayko}, as follows. We use a frequentist-based method to estimate the red noise properties of the data set \citep{Coles}\footnote{We note that the modelling could also be carried out in packages such as \textsc{enterprise}.  We chose here the frequentist approach as the signal from cusps on cosmic strings are much shorter time scale events than those currently modelled in the Bayesian analysis software.  We felt the need to inspect the power spectra of the residuals for all of the pulsars we used and to model them manually. However, we note that the results are not sensitive to the exact noise model.}.  For each pulsar, we use the \textsc{spectralmodel} plugin to look for evidence of non-white noise. We assume that red noise is the stochastic process described by the power law
\begin{equation}
P_r(f) = P_0 \left[ 1 + (f/f_c)^2 \right]^{-\frac{\gamma}{2}},
\end{equation}
where $P_0$ is the amplitude at a corner frequency $f_c$ and $\gamma$ is the power-law exponent. When such noise is present, we obtain a self-consistent estimate of the covariance matrix for the low-frequency noise using the iterative procedure discussed by \cite{Coles}. 
An initial estimate of the red noise spectrum is obtained and fitted with the power law model. The latter is used to estimate the covariance matrix of the red noise.  The white noise is modelled using the \textsc{EFAC} and \textsc{EQUAD} parameters within \textsc{tempo2} and determined using the \textsc{efacEquad} plugin. Adding the white-noise component of the variance as a diagonal matrix, we obtain the complete covariance matrix which is then used to improve our estimate of the power spectrum using a generalized least-squares fit. An improved model, including the white noise is fitted to this power spectrum and the process is iterated until a self-consistent solution is obtained. Table \ref{tab:red_noise} describes our red-noise models. In contrast to the analysis of \cite{Porayko}, we include red-noise models for PSRs~J1713$+$0747, J1732$-$5049, J1857$+$0943 and J2241$-$5236. Our red-noise models obtained by the frequentist analysis are, in most cases, consistent with the Bayesian noise analysis in \cite{Porayko} (note that the results are insensitive to slight changes to the noise modelling).

\begin{table}
	\centering
	\caption{Red noise properties for the PPTA data set used in this work. Columns 2, 3 and 4 are from this work. Columns 5, 6 and 7 are the corresponding values from Porayko et al. (2018). Note that the white noise parameters are dependent on the signal processor used and are available in the pulsar timing model data files.}
	\begin{tabular}{lp{0.1cm}p{0.3cm}lp{0.3cm}p{0.4cm}l}
		\hline
		Pulsar Name & $\gamma$ & $f_c$ & $P_0$ & $\gamma_{P18}$ & $f_{c,P18}$ & $P_{0,P18}$ \\
		    & &  $({\rm{yr}}^{-1})$ & $({\rm{yr}}^{3})$ & & $({\rm{yr}}^{-1})$ & $({\rm{yr}}^{3})$ \\
		\hline
		J0437$-$4715 & 3.5 & 0.08 & 2.66 $\times 10^{-27}$ & 3.5 & 0.08 & 2.37 $\times 10^{-27}$ \\
		J0613$-$0200 & 2 & 0.08 & 4.31 $\times 10^{-26}$ & 2.5 & 0.08 & 1.30 $\times 10^{-26}$ \\
		J0711$-$6830 & 5 & 0.08 & 2.08 $\times 10^{-25}$ & 4 & 0.08 & 3.98 $\times 10^{-26}$ \\
        J1017$-$7156 & 6 & 1.0 & 7.24 $\times 10^{-28}$ & 6 & 1.0 & 9.54 $\times 10^{-28}$ \\
        J1022$+$1001 & 2 & 0.08 & 1.66 $\times 10^{-26}$ & 2 & 0.08 & 3.04 $\times 10^{-26}$ \\
        J1024$-$0719 & 6 & 0.08 & 3.03 $\times 10^{-24}$ & 3 & 0.08 & 4.30 $\times 10^{-25}$ \\
        J1045$-$4509 & 3 & 0.3 & 7.24 $\times 10^{-26}$ & 3 & 0.3 & 7.44 $\times 10^{-27}$ \\
        J1125$-$6014 & 1 & 0.08 & 6.02 $\times 10^{-27}$ & 3 & 0.2 & 5.79 $\times 10^{-27}$ \\
        J1446$-$4701 & $-$ & $-$ & $-$ & $-$ & $-$ & $-$ \\
        J1545$-$4550 & 4 & 0.2 & 1.16 $\times 10^{-26}$ & 3 & 0.1 & 1.66 $\times 10^{-26}$ \\
        J1600$-$3053 & 2 & 0.08 & 1.19 $\times 10^{-27}$ & 2 & 0.08 & 1.05 $\times 10^{-27}$ \\
        J1603$-$7202 & 2.5 & 0.08 & 3.58 $\times 10^{-26}$ & 3 & 0.08 & 8.39 $\times 10^{-26}$ \\
        J1643$-$1224 & 2 & 0.5 & 1.04 $\times 10^{-26}$ & 1.5 & 0.08 &3.43 $\times 10^{-26}$ \\
        J1713$+$0747 & 2 & 0.5 & 6.46 $\times 10^{-29}$ & $-$ & $-$ & $-$ \\
        J1730$-$2304 & 1.5 & 0.08 & 4.93 $\times 10^{-27}$ & 2 & 0.08 & 2.17 $\times 10^{-26}$ \\
        J1732$-$5049 & 0.5 & 0.2 & 4.32 $\times 10^{-27}$ & $-$ & $-$ & $-$\\
        J1744$-$1134 & 3 & 0.4 & 9.86 $\times 10^{-28}$ & 6 & 1.0 & 2.55 $\times 10^{-28}$ \\
        J1824$-$2452A & 4 & 0.1 & 1.98 $\times 10^{-23}$ & 4 & 0.1 & 1.22 $\times 10^{-23}$ \\
        J1832$-$0836 & $-$ & $-$ & $-$ & $-$ & $-$ & $-$ \\
        J1857$+$0943 & 4 & 0.5 & 9.25 $\times 10^{-27}$ & $-$ & $-$ & $-$ \\
        J1909$-$3744 & 2.5 & 0.08 & 1.06 $\times 10^{-27}$ & 2.5 & 0.7 & 7.54 $\times 10^{-28}$ \\
        J1939$+$2134 & 4 & 0.08 & 2.42 $\times 10^{-25}$ & 4 & 0.08 & 2.50 $\times 10^{-25}$ \\
        J2124$-$3358 & 4 & 1.0 & 5.14 $\times 10^{-27}$ & 5 & 1.0 & 5.64 $\times 10^{-27}$ \\
        J2129$-$5721 & 2 & 0.2 & 4.80 $\times 10^{-27}$ & 2 & 0.08 & 1.37 $\times 10^{-26}$ \\
        J2145$-$0750 & 1.5 & 0.08 & 7.58 $\times 10^{-27}$ & 1 & 0.08 & 5.13 $\times 10^{-27}$ \\
        J2241$-$5236 & 6 & 0.8 & 1.07 $\times 10^{-28}$ & $-$ & $-$ & $-$ \\
		\hline
	\end{tabular}
	\label{tab:red_noise}
\end{table}

\section{Signal of the GW burst from a cosmic string} \label{sec:CS signal}

The waveform of the GW burst from a cusp on a cosmic string was studied by \cite{Damour}. The GW from a cusp is linearly polarized and, for a plus-polarized event, the time-domain waveform is given by
\begin{eqnarray}
 h_+(t) &=& \begin{cases}
    A_{\rm{fit}}\left[ |t - t_0|^{1/3} - \left( \frac{1}{2}W \right)^{1/3} \right] & (t_0 - \frac{1}{2}W \le t < t_0 + \frac{1}{2}W) \\
    0 & (\rm{otherwise})
  \end{cases} \\
 h_\times (t) &=& 0,
 \label{eq:waveform}
\end{eqnarray}
where $A_{\rm{fit}}$ is the amplitude, $t_0$ is the epoch when the burst peak reaches the Earth and $W$ is the duration of the burst. Cosmic string loops are considered to generate cusps efficiently ($\mathcal{O}(1)$ per oscillation period) and the GW burst is highly beamed along the direction of cusp velocity. The direction varies depending on the string configuration and is considered to be random each time \citep{Blanco-Pillado}. 
The directions of acceleration and velocity of the cosmic string are, respectively, along the polarization basis vector $\hat{m}$ and the direction of GW propagation $\hat{\Omega}$ defined below \citep{Damour:2001bk}. 
See also section~\ref{sec:discussion} for the relationship between these quantities and physical parameters. 
Note that $A_{\rm{fit}}$ has dimensionality of sec$^{-1/3}$. We define the dimensionless amplitude, which corresponds to the peak amplitude at $t=t_0$, as
\begin{equation}
A_{\rm{peak}} \equiv h_+ (t_0) = \left( \frac{1}{2} W \right)^{1/3} A_{\rm{fit}}.
\label{eq:Apeak}
\end{equation}

The timing residuals induced by GWs are given by \citet{Detweiler}:
\begin{equation}
r(t) = \sum_{a = +, \times}F^a (\hat{\Omega}, \hat{p}) \int^t \Delta h_a(\hat{\Omega}, t') dt',
\label{eq:residual}
\end{equation}
where $\hat{p}$ and $\hat{\Omega}$ are the direction of the pulsar and of the GW propagation, respectively. Here, $F^A(\hat{\Omega}, \hat{p})$ is called the antenna pattern which is given by \citet{Anholm}:
\begin{equation}
F^a(\hat{\Omega}, \hat{p}) = \frac{1}{2} \frac{\hat{p}^i\hat{p}^j}{1 + \hat{\Omega}\cdot \hat{p}}e^a_{ij}(\hat{\Omega}).
\end{equation}
The GW polarization tensors $e^a_{ij}(a = +, \times)$ are given by
\begin{eqnarray}
e^+_{ij}(\hat{\Omega}) &=& \hat{m}_i\hat{m}_j - \hat{n}_i\hat{n}_j \\
e^\times_{ij}(\hat{\Omega}) &=& \hat{m}_i\hat{n}_j + \hat{n}_i\hat{m}_j,
\end{eqnarray}
where $\hat{m}$ and $\hat{n}$ are the polarization basis vectors. In this work, we set these vectors as $\hat{m} = (\sin\alpha_s, -\cos\alpha_s, 0)$ and $\hat{n} = (\sin\delta_s\cos\alpha_s, \sin\delta_s\sin\alpha_s, -\cos\delta_s)$ with the source position in the equatorial coordinates $(\alpha_s, \delta_s)$. 
Here, $\Delta h_A(\hat{\Omega}, t)$ is the difference of the metric perturbation between the Earth and the pulsar and is given by \cite{Book}
\begin{equation}
\Delta h_a(\hat{\Omega}, t) = h_a(\hat{\Omega}, t) - h_a(\hat{\Omega}, t_p) \label{dif_metric},
\end{equation}
where $t_p = t - \tau$ with $\tau = L/c (1+\hat{\Omega}\cdot\hat{p})$ being the pulse propagation time from the pulsar to the Earth, and $L$ is the distance to the pulsar. 
The first term in Eq.\eqref{dif_metric} represents the effect of the GW on the Earth and the second term represents its effect on the pulsar. Typically the time delay between the Earth and various pulsar terms will be 100s or 1000s of years. Since the burst from a cusp on a cosmic string is highly beamed and randomly directed they will not be observed repeatedly from the same direction.

A PTA would detect either a pulsar term or an Earth term, but it would be very difficult to avoid confusing a pulsar term detection with unmodeled instrumental offsets or errors in the red noise model. Accordingly we focus on the Earth term, which not only provides multiple detections, but these detections must be correlated with the correct ``antenna pattern''.

\begin{figure}
\centering
\includegraphics[width=\linewidth]{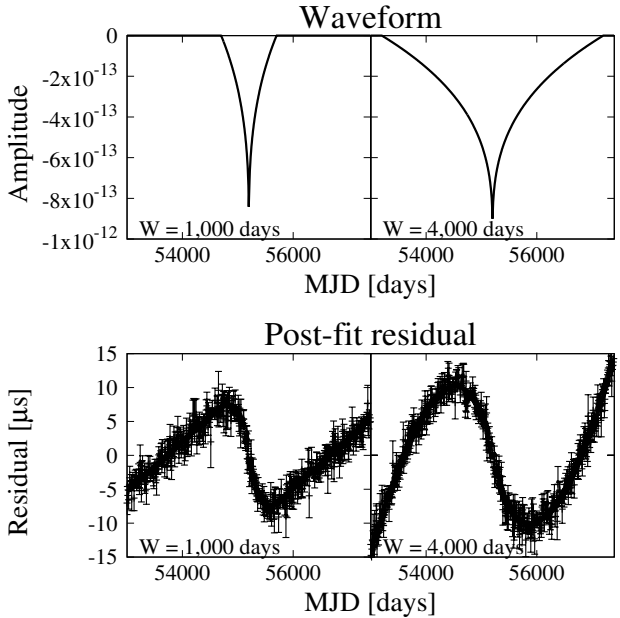}
\caption{Waveforms of the GW burst from a cosmic string cusp (top panels) and simulated post-fit timing residuals with Gaussian white noise of 1 $\mu$sec (bottom panels). Here, the injected GW amplitude is set to $A_{\rm{peak}} = 10^{-12}$ and the peak is at MJD 55200. In the bottom panel, pulsar parameters such as the pulse period and spin-down rate have been fitted and so post-fit residuals are shown.}
\label{fig:example_burst}
\end{figure}

The analytic expressions of the pre-fit timing residual induced by the GW burst from a cosmic string cusp are derived by substituting Eq.\eqref{eq:waveform} into Eq.\eqref{eq:residual} using $A_{\rm{fit}}$,
\begin{eqnarray}
  r(t) &=& F^+ (\hat{\Omega}, \hat{p}) \nonumber\\
  &\times&
  \begin{cases}
    0  & ( t < t_0 - \frac{W}{2}) \\
    \\
    A_{\rm{fit}} \left[ \frac{3}{4} \left\{ \left( \frac{W}{2} \right)^{4/3} \mp |t-t_0|^{4/3} \right\} \right . \\ 
    ~~~ \left .- \left( \frac{W}{2} \right)^{1/3} \left\{ t - \left( t_0 - \frac{W}{2} \right) \right\} \right]
    & (t_0 - \frac{W}{2} \le t < t_0 + \frac{W}{2}) \\
    \\
   -\frac{1}{4} \left( \frac{1}{2} \right)^{1/3}A_{\rm{fit}}W^{4/3} 
   & (t \ge t_0 + \frac{W}{2})
  \end{cases}\nonumber\\
   \label{eq:res_cs}
\end{eqnarray}
In the second line,  $\mp$ indicates that the $-$ sign should be applied before $t_0$ and $+$ after.
In Figure~\ref{fig:example_burst}, we show examples of the waveform and simulated timing residuals (without the antenna pattern, after fitting the pulsar parameters). The timing residuals are generated by injecting a GW signal using Eq. \eqref{eq:res_cs} with the amplitude $A_{\rm{peak}} = 10^{-12}$ at  the center of the observational span (MJD 55200) and adding Gaussian white noise of 1 $\mu$sec. The widths of the bursts are taken as 1000 (left panel) and 4000 days (right panel), which corresponds to $A_{\rm{fit}} = 2.85\times 10^{-15}$ and  $1.80\times 10^{-15}$, respectively.  Timing residuals induced by a cosmic string GW burst have different shapes depending on the epoch and the width of the burst, but they are deterministic, allowing us to perform a matched filter search.

We have incorporated the effect of the GW burst from a cosmic string into the \textsc{tempo2} timing model. This allows us to fit the GW burst from a cosmic string and to simulate residuals (or ToAs). The new timing model parameters are $(A_{\rm{fit}}, t_0, W, \alpha_s, \delta_s, \zeta)$. 
Here, $\zeta$ is the principal polarization angle.
\textsc{tempo2} uses a linear, generalised, least-squares-fitting algorithm. If the burst epoch, width, polarization angle and source position are known, we can obtain the amplitude of the GW burst as a part of the standard \textsc{tempo2} timing fit. However, if these parameters are not known, then a non-linear fitting routine is needed to determine the values.

We have found that this parametrization is convenient for simulating timing residuals caused by the GW burst from a cosmic string. However, in the search procedure, we found that it is useful to provide a second parametrization of the GW burst. In this parametrization, we describe the GW burst using two orthogonal components, $A_1$ and $A_2$ where $A_1 = A_{\rm{fit}}\cos (2\zeta)$ and $A_2 = A_{\rm{fit}}\sin (2\zeta)$. 
These $A_1$ and $A_2$ correspond to two GW polarization modes and this parametrization enables us to search for all GW polarizations. This formulation has the advantage that $A_1$ and $A_2$ enter the timing model linearly and can be fitted with linear least squares. We emphasize that, even with this parametrization, the position of the source and epoch and width of the event cannot be obtained using a linear-fitting routine. We therefore fit $A_1$ and $A_2$ for different sky locations, epochs and widths.

Users of the \textsc{tempo2} software package can employ the following parameters in their parameter files: \textsc{GWCS\_A1}, \textsc{GWCS\_A2}, \textsc{GWCS\_POSITION}, \textsc{GWCS\_EPOCH} and \textsc{GWCS\_WIDTH} for the cosmic string $A_1$, $A_2$, sky position, burst epoch and width, respectively.  The \textsc{ptaSimulate} software package to simulate PTA-style data sets can also be used to simulate cosmic string burst events as described in this paper.

\section{Detection principle}\label{sec:principle}

The timing residuals for each pulsar induced by a GW burst are determined by specifying the source parameters: $A_1$, $A_2$, position, epoch and width. For a given position, epoch and width, we jointly fit $A_1$ and $A_2$ with the pulsar parameters (spin-down, astrometry, orbital configuration, etc.) by adding the GW burst response to the \textsc{tempo2} timing model and subsequently minimizing the whitened timing residuals. To account for the correlations in the pre-fit timing residuals caused by unmodelled red noise, we use the algorithm referred to as generalized least squares described in \cite{Coles}. To determine the sky position, epoch and width, we search over a regular four-dimensional grid whose spacing we describe below.

At each position, epoch and width, \textsc{tempo2} returns the parameter vector ${\bf{A}} = [A_1;A_2]$ and the covariance matrix, ${\sf{C}}_0$, of the estimation of ${\bf{A}}$. From these we calculate a detection statistic, which provides an optimal estimate of the amplitude of the GW burst. We then use that statistic to locate the GW burst in the grid of possible positions, epochs and widths. While this approach is not as computationally efficient as a non-linear fit, it provides an opportunity to study the statistics of the noise by examining the response over the entire four-dimensional grid.

If the pulsars are uniformly distributed in the sky and have the same ToA precision then $A_1$ and $A_2$ become orthogonal (independent) and equally sensitive. For this ideal case, $D = A_1^2 + A_2^2$ is an optimal detection statistic. However, $A_1$ and $A_2$ are correlated. To construct the optimal detection statistic, we must determine ${\sf{U}}^{-1}$, which is the transformation matrix that whitens and normalizes $\bf{A}$, i.e. ${\bf{A}}_{\rm{w}} = {\sf{U}}^{-1}{\bf{A}}$. By multiplying ${\sf{U}}^{-1}$, the components of ${\bf{A}}_{\rm{w}}$ become two uncorrelated random variables with unit variance. This reduces the problem to the one for which we know that the optimal solution is $D = A_{\rm{w}1}^2 + A_{\rm{w}2}^2$. This is analogous to the way we use the Cholesky decomposition to deal with red noise in \textsc{tempo2} (see \citealt{Coles}). The solution is given by
\begin{equation}
D = {\bf{A}}_{\rm{w}}^{\rm{t}}{\bf{A}}_{\rm{w}} = {\bf{A}}^{\rm{t}}{\sf{C}}_0^{-1}{\bf{A}}\,,
\end{equation}
where the superscript t indicates the transposed matrix. In the absence of the GW burst, $D$ is the sum of the squares of two independent standard Gaussian random variables, and therefore it follows a $\chi^2$ distribution with two degrees of freedom.

We determine the detection threshold for each width since the width has a physical meaning described in section \ref{sec: physical meaning} and each detection statistic is not  independent of the width. For each width we adopt $D_{\rm{max}}$, which is the maximum $D$ over different epochs and sky positions, as our final detection statistic. The cumulative probability distribution of $D_{\rm{max}}$ is given by
\begin{equation}
    c_{D_{\rm{max}}}(D_{\rm{max}}) = \left[c_D(D_{\rm{max}})\right]^{N_{\rm{dof}}},
\end{equation}
where $c_D$ is the cumulative probability distribution of $D$ and $N_{\rm{dof}}$ is the number of degrees of freedom, or the independent samples over which the maximum is found. The probability density of $D$ is $p_D(D) = (1/2)\exp{(-D/2)}$, so that $c_D = 1 - \exp{(-D/2)}$ and 
\begin{equation}\label{eqn:cdmax}
    c_{D_{\rm{max}}}(D_{\rm{max}}) = \left[1 - \exp{(-D_{\rm{max}}/2)}\right]^{N_{\rm{dof}}}.
\end{equation}
The false-alarm probability (FAP) is $1 - c_{D_{\rm{max}}}(D_{\rm{max}})$.

Here, the degree of freedom $N_{\rm{dof}}$ is expressed as the product of the degrees of freedom of the sky positions and epochs. Each epoch is completely independent for $W < 250$ days, which is the chosen spacing of the epochs, while they are dependent for a larger width. Thus, the degree of freedom is given by
\begin{equation}
    N_{\rm{dof}} = N_{\rm{dof, sky}}\times
    \begin{cases}
    N_{\rm{e}} & (W < 250~{\rm{days}}) \\
    \frac{4000~{\rm{days}}}{W} & (W \ge 250~{\rm{days}}),
    \end{cases} \\
    \label{eq:Ndof}
\end{equation}
where $N_{\rm{dof, sky}}$ is the number of independent samples in the entire sky and $N_{\rm{e}} = 17$ is the number of epochs searched, and 4000 days corresponds to the observational time span of the PPTA data set. We do not know $N_{\rm{dof, sky}}$, and therefore carried out a simulation to determine it and consequently the FAP.

\begin{figure}
\centering
\includegraphics[width=55mm, angle=-90]{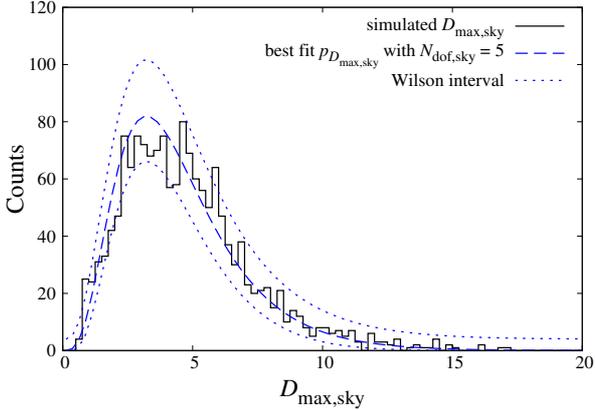}
\caption{Histogram of simulated $D_{\rm{max,sky}}$ and the best fit probability density of $D_{\rm{max,sky}}$ to the simulation data. Dotted lines show the Wilson score interval.}
\label{fig:Dsim_hist}
\end{figure}

We simulated 100 realizations of the PPTA data set.  We used the same sampling, observing frequencies and ToA uncertainties as in the real data and added in red noise at the level as defined in Table~\ref{tab:red_noise}. No GW signal was included in the simulation.  We then calculated $D_{\rm {max,sky}}$ for each realization determined over 1034 sky positions for eight epochs with a 500 day spacing and two widths (50 and 100 days) which means that the number of $D_{\rm{max,sky}}$ values in the simulation is 1600. We found the probability density of $D_{\rm{max,sky}}$, $p_{D_{\rm max,sky}}$,  by differentiating $c_{D_{\rm{max,sky}}} = \left[c_D(D_{\rm{max}})\right]^{N_{\rm{dof,sky}}}$ and fitted to the histogram of the simulations. This gave the best estimate of $N_{\rm{dof,sky}}$ = 5. The histogram and the best fit $p_{D_{\rm{max,sky}}}$ are shown in Figure~\ref{fig:Dsim_hist}.   Here we weighted the fit by the histogram error, which has a binomial distribution. In this Figure we overplot the Wilson score interval, which shows that the data are well matched with this model. The detection threshold $D_{\rm th}$  for a given FAP can then be written as
\begin{equation}
    D_{\rm th} = -2 \ln \left(1-[1-{\rm FAP}]^{1/N_{\rm dof}}\right).
\end{equation}

$D_{\rm th}$, which is a function of $W$, is plotted for a FAP = 1\% on Figure~\ref{fig:DS_obs} as a dotted line. It increases for smaller widths because $D$ is maximized over more independent trials (i.e. $N_{\rm dof}$ increases with smaller $W$).

\section{Application to the PPTA Data Set} \label{sec:results}

\subsection{Apparent detection events}

To apply our detection algorithm to the PPTA data set we first performed global fitting for $A_1$ and $A_2$ at each sky position, epoch and width, while simultaneously fitting the parameters specific to each pulsar. We obtain the detection statistic $D$ for 1034 sky positions, 17 epochs from MJD 53250 to MJD 57250 at intervals of 250 days and 13 widths (50, 75, 100, 150, 200, 300, 500, 750, 1000, 1500, 2000, 3000 and 4000 days).  A histogram of $D$ is shown in Figure~\ref{fig:Dobs_hist} as a black solid line. The histogram has been normalized to the scale of a probability distribution. A blue solid line shows the expected exponential distribution. Clearly the observations include a signal which is not consistent with the noise models. This could be caused by the GW burst event that we are searching for, a receiver configuration change that is not properly calibrated, or some other unknown event.	

\begin{figure}
\centering
\includegraphics[width=55mm, angle=-90]{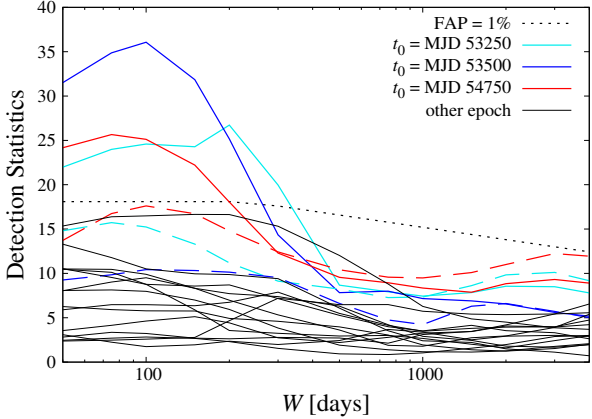}
\caption{Maximum detection statistics over the sky obtained from the PPTA data set against widths for 17 epochs. Black dotted line shows the detection threshold of $D_{\rm{max}}$ with the FAP = 1\%. Solid and dashed lines represent the original and corrected $D_{\rm{max,sky}}$ for each epoch.}
\label{fig:DS_obs}
\end{figure}

\begin{figure}
\centering
\includegraphics[width=55mm, angle=-90]{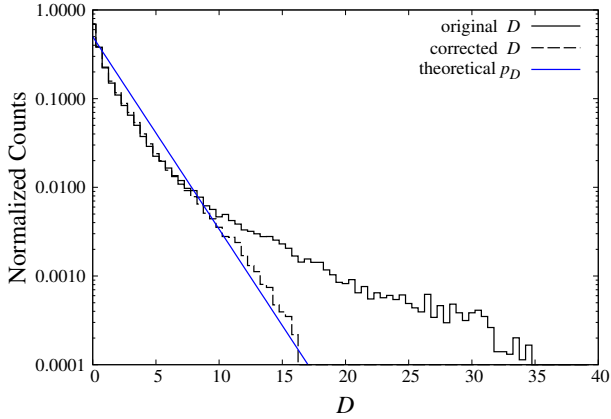}
\caption{Normalized histograms of observed $D$ and the theoretical probability density of $D$. Solid and dashed histograms show the original and after the events described in the text have been removed, respectively.}
\label{fig:Dobs_hist}
\end{figure}

\begin{figure}
\centering
\includegraphics[width=50mm, angle=-90]{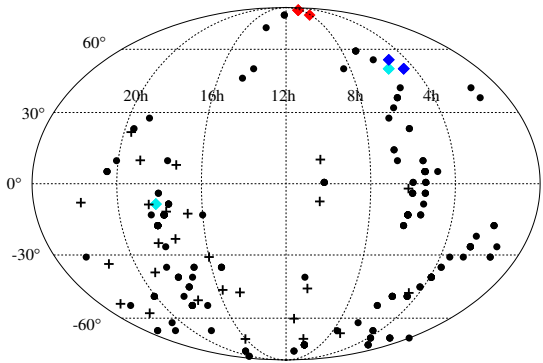}
\caption{Locations of all $D_{\rm{max,sky}}$ and the PPTA pulsars described as black dots and black crosses in the equatorial coordinates. Colored squares describes $D_{\rm{max,sky}}$ of greater than 18. Cyan, blue and red ones represent the epochs of MJD 53250, 53500 and 54750, respectively.}
\label{fig:Dmax_pos}
\end{figure}

\begin{figure}
\centering
\includegraphics[width=65mm,angle=-90]{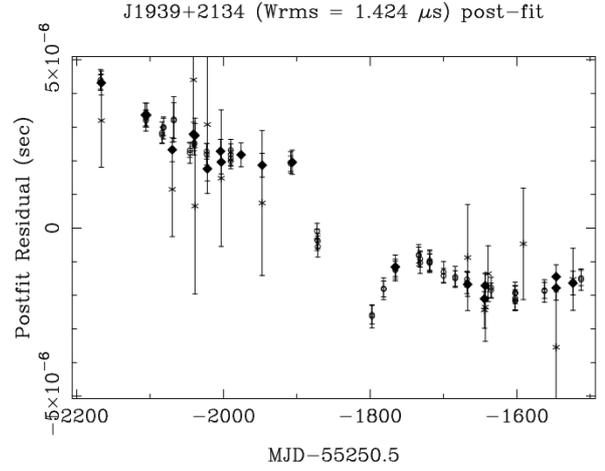}
\caption{Timing residuals for PSR~J1939+2134 between MJD~53250 and 53500. A bump-like structure can be seen that mimics a cosmic string burst event.}
\label{fig:bump}
\end{figure}

We searched for the epoch at which these apparent detection events occurred by plotting the maximum of $D$ over the sky ($D_{\rm max}$) versus width, separately for each of the 17 epochs. These are over-plotted on Figure~\ref{fig:DS_obs} as solid lines. The values of $D_{\rm max}$ in excess of the detection threshold comes from MJDs 53250, 53500 and 54750, for which the lines are in colour. The sky locations for each case of $D_{\rm max} > 18$ are shown in Figure~\ref{fig:Dmax_pos}. The apparent detections at MJDs~53500 and 54750 come from one sky location (each), but the apparent detection at MJD~53250 appears to come from two different sky locations approximately antipodal. So there are actually four apparent detections. Examining the timing residuals for each pulsar in the vicinity of epochs MJD 53250 to 53500 showed significant anomalies in the residuals for PSR~J1939+2134 between MJD 53400 and 53460. These are shown in Figure~\ref{fig:bump}. Accordingly we reanalyzed the observations without PSR J1939+2134 and found that the high values of $D_{\rm max}$ were eliminated both at MJDs 53250 and 53500.
These events could be detections of the pulsar term, but detection of three pulsar term bursts at three different sky locations, in the same pulsar, stretches credulity. This is particularly true when there is a big data-gap in the observations of that pulsar. Accordingly we 
conclude that the events of MJD~53250 and 53500 are spurious.

The removal of PSR~J1939+2134 did not eliminate the apparent detection at MJD 54750, so we searched for a possible instrumental cause. We found that there was a receiver configuration change at MJD 54751 (11th Oct. 2008). At this time, an update was made to our primary observing signal processing system (the PDFB4) and it is likely that a phase offset at this time was not entirely accounted for. This would have affected several pulsars and we found that removing both PSR~J1939+2134 and J0437$-$4715 eliminated this apparent detection.  These two pulsars have very low white noise, so they are heavily weighted in the detection algorithm, but they also have high red noise and are among the pulsars most difficult to model. Accordingly we do not claim a detection at MJD~54750, but we note the parameters of this apparent burst in Table~\ref{tab:suprious}.  

\begin{table}
	\centering
	\caption{Measured parameters of the event in excess of the detection threshold at MJD 54750.}
	\begin{tabular}{ccccc}
		\hline
        $W$ [days] & $\alpha_s~[^\circ]$ & $\delta_s~[^\circ]$ & $A_1$ & $A_2$ \\
		\hline
		 50 & 122.8 & 81.6 & $8.88\times 10^{-15}$ & $8.04\times 10^{-15}$ \\
		 75 & 131.4 & 86.1 & $6.83\times 10^{-15}$ & $3.16\times 10^{-15}$ \\
		 100 & 131.4 & 86.1 & $5.06\times 10^{-15}$ & $2.07\times 10^{-15}$ \\
		 150 & 131.4 & 86.1 & $3.36\times 10^{-15}$ & $9.74\times 10^{-16}$ \\
		\hline
	\end{tabular}
	\label{tab:suprious}
\end{table}

We have recomputed the histogram of $D$ for all sky positions, epochs, and widths, except that we have now excluded PSR~J1939+2134 when fitting for epochs 53250, 53500 and 54750, and also PSR~J0437$-$4715 when fitting for epoch 54750. This histogram is plotted as a dashed line in Figure~\ref{fig:Dobs_hist}. Clearly the detection statistics are now consistent with the expected exponential distribution. We also over-plotted the $D_{\rm max}$ for epochs 53250, 53500, and 54750 as dashed lines on Figure~\ref{fig:DS_obs}, and the results show that, with this processing, no GW detection is made with a FAP of 1\%.
	
\subsection{The sensitivity of the data set}

Using the value of $D_{\rm{max,sky}}$ and the covariance matrix, we can determine the sensitivity of the data set to the GW burst for each epoch, width and sky position. For a GW burst with amplitude $h^{\rm{CS, fit}}$ with random polarization, the expected values are $\langle A_1^2 \rangle  = \langle A_2^2 \rangle = (h^{\rm{CS, fit}})^2/2$ and $\langle A_1 A_2 \rangle = 0$, and we obtain
\begin{equation}
\langle D \rangle = \left( h^{\rm{CS, fit}} \right)^2 \left( S_{11} + S_{22} \right)/2,
\end{equation}
where ${\sf{S}} = {\sf{C}}^{-1}_0$ and ${\sf{C}}^{-1}_0$ can be obtained directly from fitting the PPTA data. $S_{11}$ and $S_{22}$ are the diagonal elements of $\sf{S}$. Thus, the GW burst amplitude corresponding to a $D_{\rm{max,sky}}$ for a given epoch and width is
\begin{equation}
h^{\rm{CS, fit}} = \left(2D_{\rm{max,sky}}/[S_{11} + S_{22}] \right)^{1/2}. 
\label{eq:sensitivity}
\end{equation}
Here, $h^{\rm{CS, fit}}$ corresponds to $A_{\rm fit}$ in Eq. \eqref{eq:waveform} when the GW is plus-polarized. 
As mentioned in section \ref{sec:CS signal}, this has a dimension of sec$^{-1/3}$, and the dimensionless GW amplitude is given by 
\begin{equation}
h^{\rm{CS, peak}} = \left( \frac{1}{2}W \right)^{1/3} h^{\rm{CS, fit}}.
\end{equation}

\begin{figure}
\centering
\includegraphics[width=8.5cm]{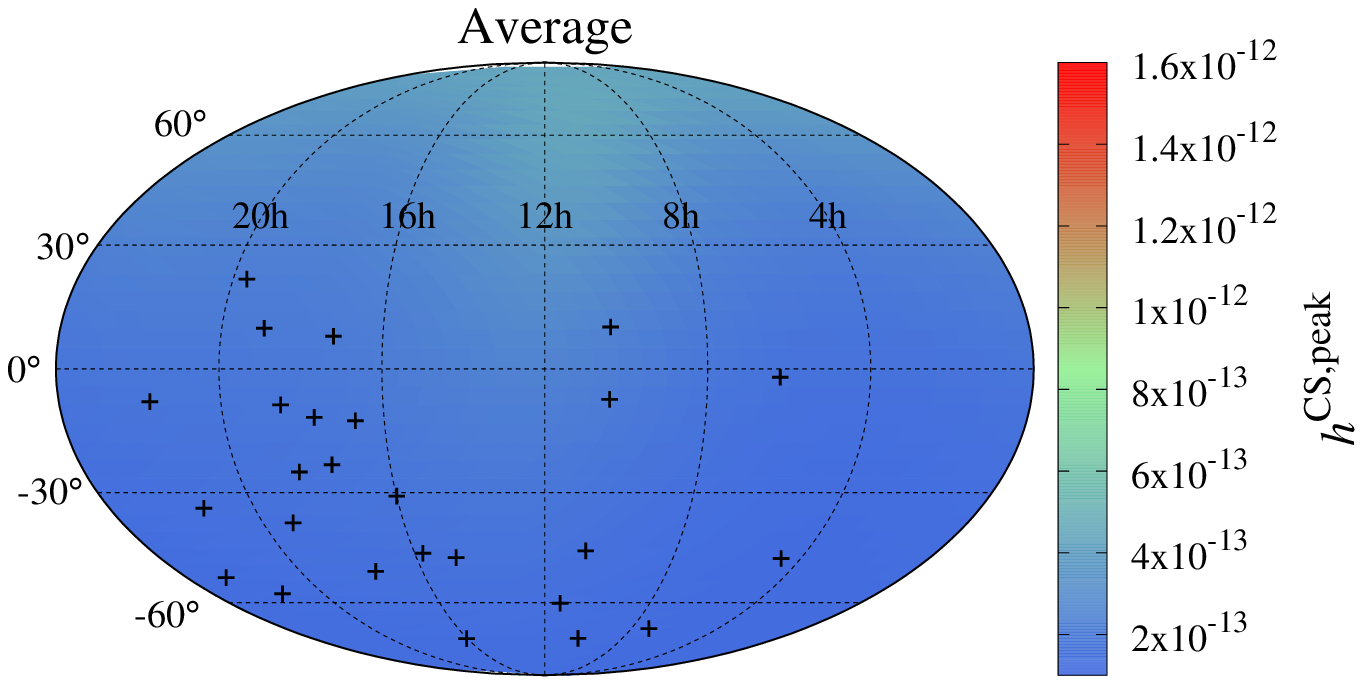}
\includegraphics[width=8.5cm]{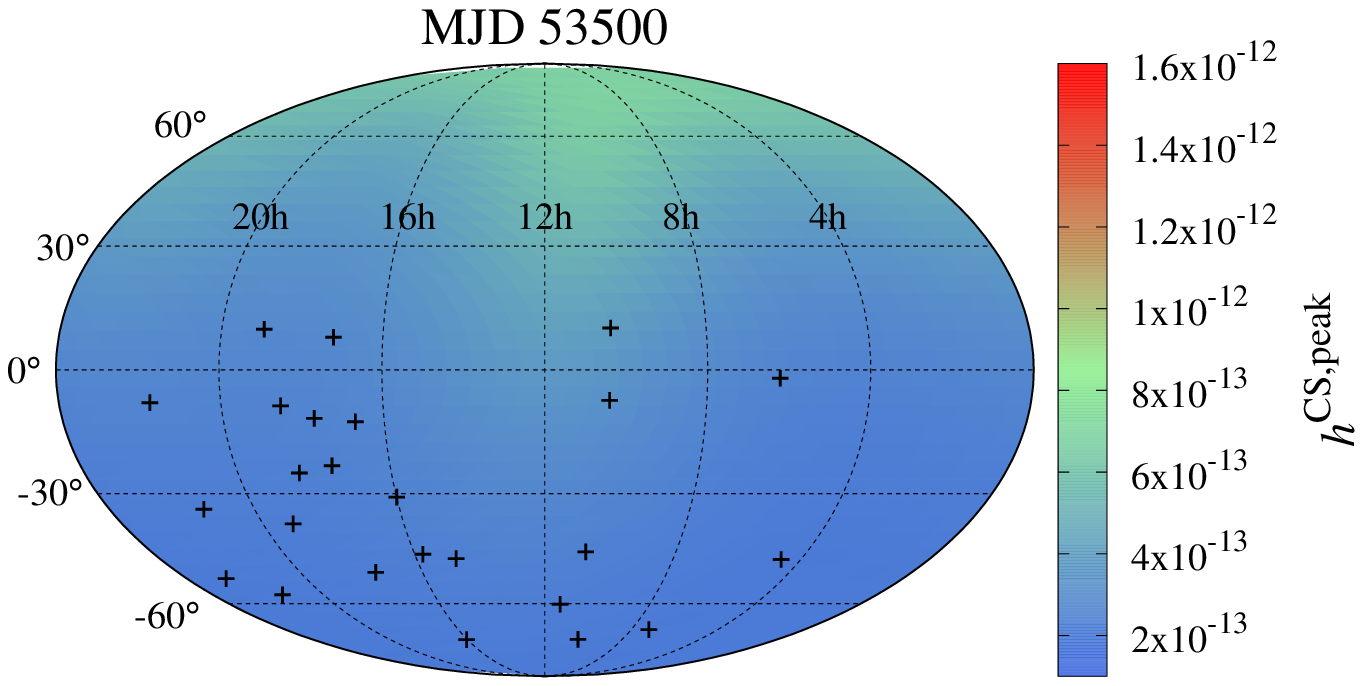}
\includegraphics[width=8.5cm]{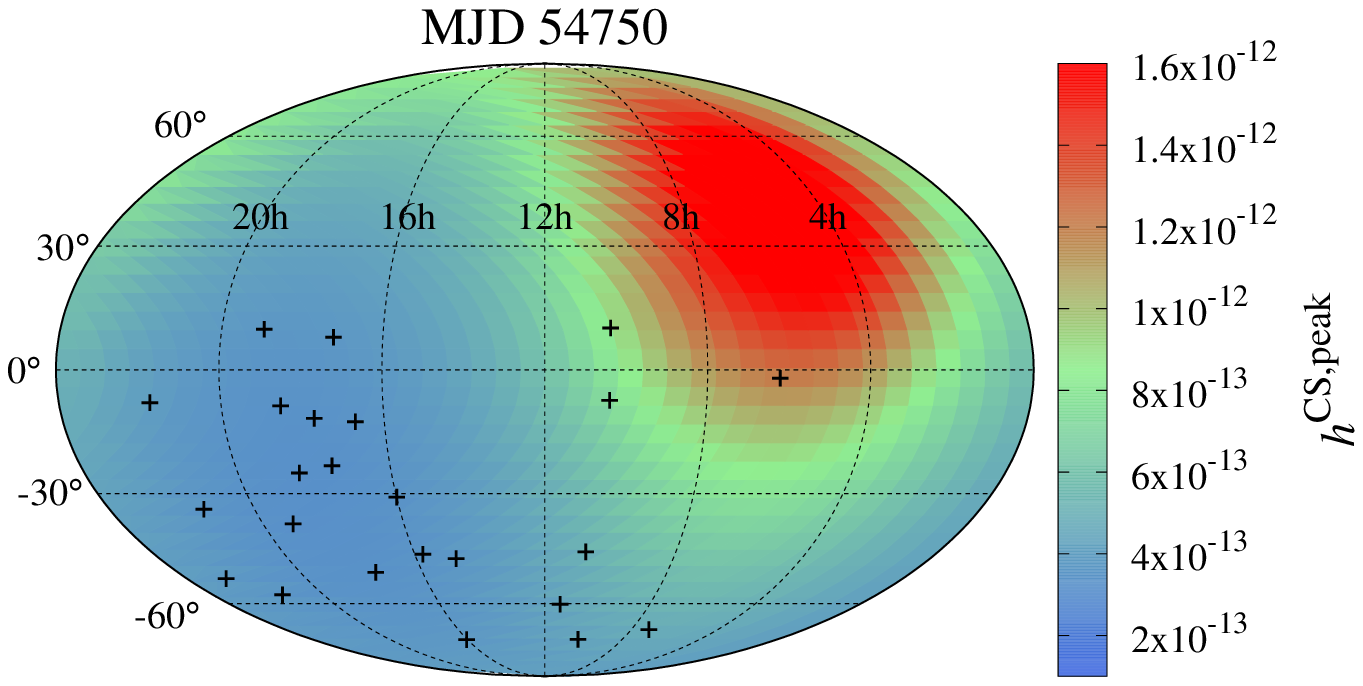}
\caption{Sensitivity of the PPTA data set for $h^{\rm{CS, peak}}$ for a cosmic string width of 100\,d in equatorial coordinates.  Black crosses represent the positions of the PPTA pulsars.  The top panel includes all pulsars and the sensitivity is measured over all epochs apart from MJD~54750. In the central panel PSR~J1939$+$2134 has been removed and the sensitivity determined for MJD~53500.  In the bottom panel both PSRs~J0437$-$4715 and J1939+2134 are removed and the sensitivity determined for MJD~54750.}
\label{fig:sensitivity}
\end{figure}

In order to obtain sensitivity maps the covariance matrix is calculated at each sky position and is converted to the value of $h^{\rm{CS, fit}}$ using Eq. \eqref{eq:sensitivity} with the false alarm probability of 1\% for a width of 100\,d ($D_{\rm max,sky}= 17.6$).  Using these calculations we find that the sensitivity is significantly worse at MJD 54750 even though we have removed PSRs~J0437$-$4715 and J1939+2134 from the data set for that epoch (see bottom panel of Figure~\ref{fig:sensitivity}).  As described earlier this is explained as an instrumental system reconfiguration that likely affected all pulsars. Accordingly we removed the epoch MJD~54750 from subsequent processing. The top panel of Figure~\ref{fig:sensitivity} shows the sensitivity map averaged over all epochs apart from MJD~54750. This shows that the PPTA pulsar timing array is slightly more sensitive for a GW burst in the Southern area as pulsars are concentrated in the Southern hemisphere.

The removal of PSR~J1939+2134 from epochs MJDs~53250 and 53500 has little effect on the sensitivity, as shown in the central panel of Figure~\ref{fig:sensitivity}.  As this is a Northern pulsar, its removal slightly degrades the sensitivity in the North.  The bottom panel of the Figure clearly shows that the additional removal of PSR~J0437$-$4715 significantly reduces the sky sensitivity, in particular in the direction around that pulsar.  This is not unexpected as that pulsar, when present, contributes significant weight to the fitting procedures because of the large number of observations and high time precision.

\begin{figure}
\centering
\includegraphics[width=8cm]{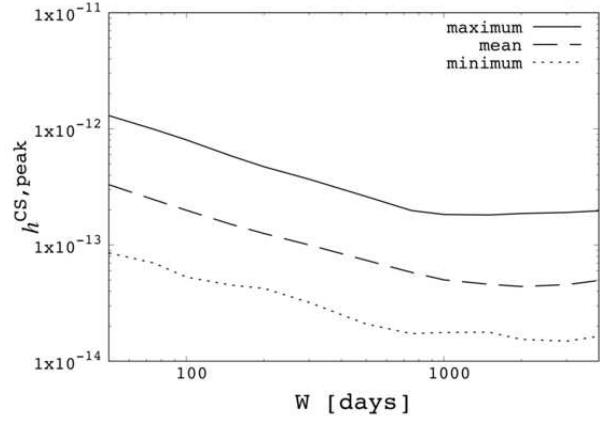}
\caption{Constraint on the GW amplitude as a function of the width $W$. The maximum and minimum values of $h^{\rm{CS, peak}}$ are determined by searching over epochs and sky positions for a fixed event width.}
\label{fig:h_peak}
\end{figure}

The constraints on $h^{\rm{CS, peak}}$ can be provided as a function of the epoch, width and sky position. Since only the width has a physical meaning, which corresponds to the loop size of the cosmic string, in Figure~\ref{fig:h_peak}, we show the maximum value of $h^{\rm{CS, peak}}$ as a function of the width, searched over different epochs and sky positions.   We excluded MJD~54750 when determining these maximum values. However, we kept in MJDs~53250 and 53500 because, as discussed earlier, the removal of PSR~J1939+2134 alone did not significantly increase the noise level. We find that the constraint on $h^{\rm{CS, peak}}$ gets stronger as $W$ increases. The reason is that timing residuals induced by the GW burst become larger for a longer event, since a timing residual is given by the time integration of $\Delta h$.

\section{Discussion} \label{sec:discussion}

In this section we discuss the physical implication of the above results. First we describe how the waveform given in Eq. (3) corresponds to the cosmic string parameters. Next, using the theoretical estimate on the event rate given in the Appendix we translate the upper limit on the GW amplitude obtained in the previous section to the constraint on the cosmic string tension. We then provide implications for observations by the IPTA and the SKA.

\subsection{Physical meaning of the waveform} \label{sec: physical meaning}
The spectral shape of GWs from a cusp was first investigated in \cite{Damour} and found to follow a simple power law $\propto |f|^{-4/3}$. The strain amplitude in the frequency domain $\tilde{h}(f)=\int dt e^{2\pi ift}h(t)$ is given in terms of the string tension $G\mu$ and the loop size $l$ as 
\begin{equation}
  \tilde{h}(f) =  \frac{G\mu \cdot l}{[(1+z)f \cdot l]^{1/3}r(z)f} \,,
  \label{eq:amp}
\end{equation}
where $r(z)$ is the comoving distance to the GW source at redshift $z$. 
Here the string tension is described as a dimensionless parameter $G\mu$, where $\mu$ is the energy per unit length stored in a cosmic string and $G$ is the gravitational constant. For a field theoretical string, the tension value is close to the square of the energy scale of the spontaneously super-symmetry breaking. 

The string loop cannot emit GWs of wavelength larger than its size $l$. Thus, the spectrum peaks at $f = f_l\equiv (l/2)^{-1}$ and is cut off below this frequency. This lowest frequency, determined by the loop size, also gives the duration of the GW event as $T_l = l/2 = f_l^{-1} \equiv W/2$.  Thus, searching bursts for different durations $W$ corresponds to searching for cosmic string loops of different loop size.

Let us define 
\begin{equation}
    \tilde{A}_f\equiv \frac{ G\mu\cdot l^{2/3}}{(1+z)^{1/3}r(z)}
\end{equation} 
so that the Fourier amplitude can be written in the simple form $h(f) = \tilde{A}_f|f|^{-4/3}$. 
Then, taking the inverse Fourier transform, we find that the strain amplitude in the time domain is
\begin{equation}
  h(t) = \sqrt{\frac{3}{2\pi}}\Gamma({\textstyle -\frac{1}{3}}) \tilde{A}_f |t|^{1/3} \,.
  \label{eq:ht}
\end{equation}
The coefficient $\sqrt{3/(2\pi)}\Gamma({\textstyle -1/3}) \tilde{A}_f$ corresponds to $A_{\rm fit}$ in Eq. \eqref{eq:waveform}. This function implies $h(t=0)=0$, but the offset of the amplitude can change since the full waveform is the sum of Eq. \eqref{eq:ht} and a slowly varying component due to the low modes of the string. However, only the relative difference $\Delta h$ is important for PTA observations, and the offset does not matter. In fact, we add the offset $(W/2)^{1/3}A_{\rm fit}$ in Eq. \eqref{eq:waveform} in order to set $h(t=t_0\pm W/2)=0$, but it is absorbed when we fit the pulse frequency and it does not affect the post-fit residuals. 

The important feature is that the function is spiky at $t=0$. (In Eq. \eqref{eq:waveform}, we generalize the position of the spike by adding $t_0$.) This sharp spike in $h(t)$ leaves the feature in the timing residual as shown in Figure~\ref{fig:example_burst}. Finally, considering that the time scale of the GW event is given by $T_l = f_l^{-1}$, the peak amplitude, $\Delta h$, can be written in terms of the Fourier amplitude as 
\begin{equation}
  A_{\rm peak} \sim \sqrt{\frac{3}{2\pi}}\Gamma({\textstyle -\frac{1}{3}}) \tilde{A}_f T_l^{1/3}
  = \sqrt{\frac{3}{2\pi}}\Gamma({\textstyle -\frac{1}{3}})\tilde{h}(f_l)f_l \,.
  \label{eq:Apeak_theory}
\end{equation}
This corresponds the dimensionless amplitude in Eq. \eqref{eq:Apeak}. 

Note that Eq. \eqref{eq:amp} has a high frequency cut-off if the observer does not lie exactly along the direction of the cusp velocity. This cut-off rounds off the spike in the time-domain waveform and reduces the amplitude of the timing residual. Since the timing residual is the cumulative of $\Delta h$ as in Eq. \eqref{eq:residual}, the shape of the timing residual is not changed much except for the amplitude. Thus, it does not affect the analysis performed in the previous section, although the effect on the amplitude should be taken into account when we consider constraints on string parameters. 

Here we roughly estimate how much the timing residual is reduced. According to \cite{Damour:2001bk}, the spike is smoothed by a time interval of order $|t-t_0|\sim \theta^3 T_l$, where $\theta$ is the angle (in radians) between the direction of emission and the cusp velocity. Thus, the peak amplitude is reduced by a factor of $(1-\theta^3)^{1/3}$. In the Fourier space, it means that modes with frequencies higher than $|f| \sim (\theta^3 T_l)^{-1}$ exponentially decay. Here, we find that, for $\theta\gtrsim 1$, even the lowest frequency $f_l=T_l^{-1}$ is smoothed out, so that the maximum angle we can observe the GW is $\theta\sim 1$. In other words, the GW burst from a cusp is highly beamed in the angle of $\theta < 1$. Taking an average in the solid angle of $\Omega = 2\pi (1-\cos[1])$, we find that the typical smoothing time scale is
\begin{equation}
  |t-t_0| = \frac{T_l}{\Omega}\int^{2\pi}_0 d\phi\int^1_0 d\theta \sin\theta \cdot \theta^3\simeq 0.39T_l \simeq 0.2 W  
\end{equation}
and the amplitude is reduced by a factor of
\begin{equation}
  \beta=\frac{1}{\Omega}\int^{2\pi}_0 d\phi\int^1_0 d\theta \sin\theta (1-\theta^3)^{1/3}\simeq 0.81.  
\end{equation}

One may think that we can observe a GW burst with a width larger than the observation time, $W > T_{\rm obs}\simeq 4400~{\rm days} ~ (12~{\rm years})$, if the spiky shape lies within the observation period. However, such bursts are typically smoothed out by this high-frequency cut-off as the spiky feature is rounded off for $|t-t_0| < 0.2 W$, which means that we do not see the typical cusp-origin GW shape in the range of observation period if $W>5T_{\rm obs}$. 

\subsection{Constraint on the cosmic string tension}

Here we translate the upper limit on the peak amplitude shown in
Figure~\ref{fig:h_peak} to the constraint on the cosmic string tension.  
As derived in the Appendix, the event rate of GW bursts from cosmic string cusps emitted in the redshift between $z$ and $z+dz$ whose Fourier amplitude is between $\tilde{h}$ and $\tilde{h}+d\tilde{h}$ can be written using various parameters as  
\begin{eqnarray}
\frac{dR}{dzd\tilde{h}} 
& = & \frac{3}{4}\theta_m(f,z,l)^2\frac{N_c}{(1+z)\tilde{h}}\frac{C}{\alpha p^2} \frac{1}{(\alpha + \Gamma G\mu)t_b^4}
\nonumber\\
& ~ & \times\left(\frac{a(t_b)}{a(t)}\right)^3 \frac{dV}{dz}\Theta(f \cdot l(1+z)-2) \,.
\label{eq:dRdzdh1}
\end{eqnarray}
where $\theta_m(f,z,l)\equiv [(1+z)f \cdot l]^{-1/3}$, $N_c$ is the number of cusps per oscillation period of the string loop $T_l=l/2$, $C$ is the parameter characterizing the number of string loops, $\alpha$ characterises the loop size at formation, $p$ is the probability of recconection when two strings intersect, $\Gamma$ represents the efficiency of GW emission from the loop, $t_b$ is the time when the loop is formed and given by 
$t_b = (l + \Gamma G\mu \cdot t)/(\alpha + \Gamma G\mu)$. 
In the following analysis, we use the values commonly used in the literature, $N_c=1$, $p=1$, $C_r=13.7$ ($C$ for the radiation-dominated era), $C_m=2.63$ ($C$ for the matter-dominated era), and $\Gamma=50$. Recent simulations (\cite{Ringeval, Blanco-Pillado2}) suggest the large initial loop size of $\alpha\sim 0.1$, while it could be much smaller $\alpha\sim \Gamma G\mu$ if the gravitational backreaction accounts for the loop formation. Thus, we investigate the constraints on tension for different values of $\alpha$.
In the second line, $a(t)$ or $a(z)=[1+z]^{-1}$ is the scale factor of the expanding Universe and the volume of the Universe is given as 
\begin{equation}
\frac{dV}{dz}=\frac{4\pi a(z)^2 r(z)^2}{H(z)(1+z)}
\end{equation} 
where $H(z)$ is the Hubble expansion rate and given by $H(z)=H_0\sqrt{\Omega_r (1+z)^4+\Omega_m (1+z)^3+\Omega_\Lambda}$ with $H_0=100h$~km/s/Mpc. They can be calculated once we set the cosmological parameters and we use $h=0.678$, $\Omega_m=0.308$, $\Omega_r h^2=4.15\times 10^{-5}$, and $\Omega_\Lambda=1-\Omega_m$ \citep{Ade2016}. The Heaviside step function $\Theta$ represents a low-frequency cut-off below $f_l=(l/2)^{-1}$ at the emission.

Then the number of GWs coming to the Earth today above the strain amplitude $\tilde{h}_{\rm lim}$ is given by
\begin{equation}
  N_{\rm GW} = T_{\rm obs}\int_{\tilde{h}_{\rm lim}}^{\infty} d\tilde{h} \int_0^{\infty} dz \frac{dR}{dzd\tilde{h}},
  \label{eq:NGW}
\end{equation}
where we take the observation time to be $T_{\rm obs}=12$~years.
Considering no detection of cosmic string GW bursts with amplitude above $\tilde{h}_{\rm lim}$, the parameter space giving $N_{\rm GW}>2.996$ is excluded at the $95\%$ level for a random Poisson process. We have provided upper bounds on the time-domain peak amplitude $h^{\rm CS,peak}$ for different widths $W$ in Figure~\ref{fig:h_peak}. This can be translated to the Fourier strain amplitude $\tilde{h}_{\rm lim}$, using Eq. \eqref{eq:Apeak_theory}, as
\begin{equation}
  \tilde{h}_{\rm lim}(f_l) \sim \frac{h^{\rm CS,peak}(W=2f_l^{-1})}{\beta \sqrt{\frac{3}{2\pi}}\Gamma({\textstyle -\frac{1}{3}})f_l }\,,
  \label{eq:hlim}
\end{equation}
where we have added the suppression factor $\beta=0.81$ in order to take into account the effect of the high-frequency cut-off. In the following, we place an upper limit on the cosmic string tension $G\mu$ using Eqs.\eqref{eq:NGW} and \eqref{eq:hlim}. 
Note that Eq.\eqref{eq:NGW} is obtained assuming that sensitivity is uniform over the sky ($dV/dz$ has a factor of $4\pi$ which comes from the full sky integration) and also over the observing time $T_{\rm obs}$. However, in reality the pulsar timing has a non-uniform sensitivity in both respects as can be seen in Figure \ref{fig:sensitivity}. In computing the bound on $G\mu$, we use the maximum value of $h^{\rm CS,peak}$ found over epochs and sky positions (black solid line in Figure~\ref{fig:h_peak}), so that our bound is conservative. 

The maximum of $h^{\rm CS,peak}$ shown in Figure 8, is nearly an order of magnitude greater than the mean, so there is considerable scope for reducing the upper bound on $G\mu$. We estimate that reducing the sky coverage by 10\% to exclude that portion of the sky with the highest $h^{\rm CS,peak}$ could improve our upper bound on $G\mu$ by about half an order of magnitude. A more optimal approach would be to integrate the event rate over the entire sky and observing time. However this would demand a more sophisticated analysis, which does not appear to be justified by the available data.

\begin{figure}
  \begin{center}
  \includegraphics[width=9cm]{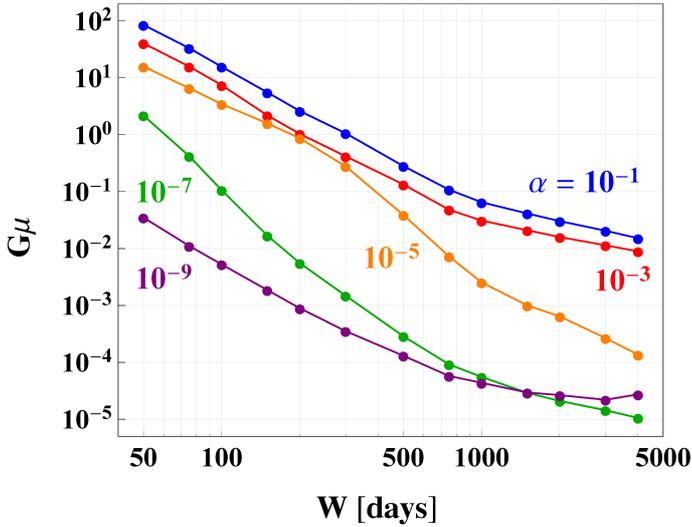}
  \caption{Constraint on $G\mu$ obtained for different widths $W$.  We show the cases for different initial loop sizes $\alpha = 10^{-1},10^{-3},10^{-5},10^{-7},10^{-9}$. Here we assume $p=1$.}
  \label{Fig:constraint1}
  \end{center}
\end{figure}

Figure \ref{Fig:constraint1} shows the constraint on the string tension $G\mu$ for different initial loop sizes $\alpha$, obtained using $h^{\rm CS,peak}$ for different widths $W$. We see that the constraint improves for larger $W$ for the following two reasons. First, the bound on $h^{\rm CS,peak}$ is more constraining for larger $W$, as shown in Figure~\ref{fig:h_peak}. Second, loops with larger size have a larger GW amplitude and a smaller number density. The former makes detection easier and the latter makes it more difficult. Taking into account both effects, we find that the former effect dominates and larger loops, equivalently larger $W$, are easier to detect and give better constraints. We also see that the constraint is stronger for smaller initial loop sizes. 
The lifetime of the loop is given by $(\alpha t_b)/(\Gamma G\mu)$. This means that loops live longer when $\alpha$ is large, so that the loops we observe today are generated in the earlier stage of the Universe, i.e. the radiation-dominated era. On the other hand, in the case of small $\alpha$, loops existing today are formed in the matter-dominated era, since loops generated in the radiation-dominated era do not survive until today because of their short lifetime. It is known that a larger number of loops are created in the matter-dominated era compared to the radiation-dominated era because of the difference in the expansion rate. Thus, since we expect more loops from the matter-dominated era when $\alpha$ is small, the constraint gets stronger for smaller $\alpha$. 

\begin{figure}
  \begin{center}
  \includegraphics[width=9cm]{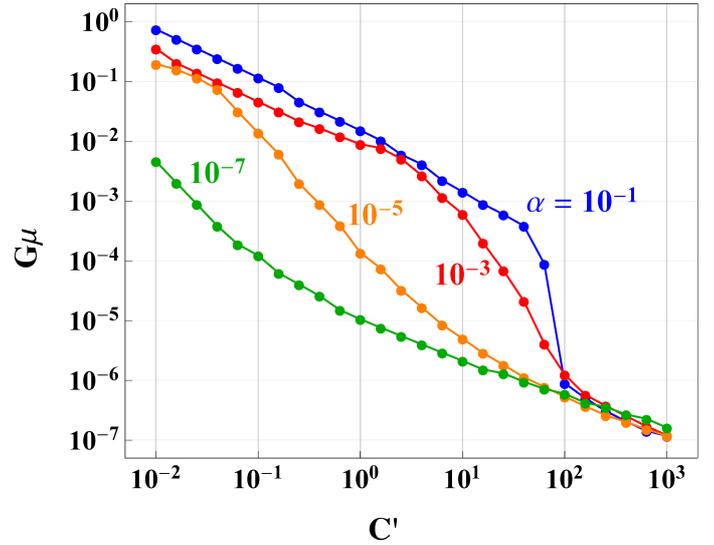}
  \caption{Dependence of the upper bound of $G\mu$ on the coefficient of loop number density $C'$ for $W=4000$~days.}
  \label{Fig:constraint2}
  \end{center}
\end{figure}

The parameters such as $N_c$, $C$, and $p^2$ in Eq. \eqref{eq:dRdzdh1}
change the overall number of GW bursts and affect the upper bound.
The number of cusps per oscillation period $N_c$ is typically
considered to be of order $1$.  The coefficient in the loop number
density $C$ is determined by the number of long strings inside the horizon and by the efficiency of energy loss to loops.  This value can vary in the range ${\mathcal O}(1-10)$. For example, we use $C_r=13.7$ and $C_m=2.63$, while the LIGO paper \citep{Abbott2018} uses $C_r=1.6$ and $C_m=0.48$ (for Model 1). Lastly, the reconnection probability $p$ can vary a lot, as it may get suppressed up to ${\mathcal O}(10^{-3})$ in the case of superstrings \citep{Jackson:2004zg}, while field theoretic strings have $p\sim 1$.  In order to see how the constraints are affected by those factors, we define the combination of the parameters as\footnote{Here, we assume that $C_r$ and $C_m$ are changed by the same factor.}
\begin{equation}
  C' = \left(\frac{N_c}{1}\right)
  \left(\frac{C_r}{13.7}\right)
  \left(\frac{1}{p^2}\right)
       ~~~{\rm or}~~~
  \left(\frac{N_c}{1}\right)
  \left(\frac{C_m}{2.63}\right)
  \left(\frac{1}{p^2}\right)
\end{equation}
and, in Figure~\ref{Fig:constraint2}, we show how the upper bound on $G\mu$ changes when $C'$ has different values by fixing $W=4000$~days, which gives the strongest constraint in Figure~\ref{Fig:constraint1}.  We find that the upper bound on $G\mu$ gets better for larger $C'$ as it becomes easier to detect GWs when the number density increases. The jump in the curve, most prominent for $\alpha=10^{-1}$, corresponds to whether the loops are produced in the radiation or matter era.

\begin{figure}
  \begin{center}
  \includegraphics[width=9cm]{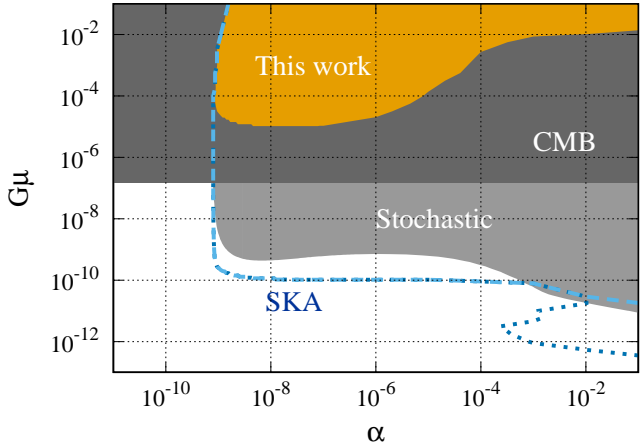}
  \caption{Constraint in the $G\mu$ --$\alpha$ plane obtained by this work (orange) compared with the constraint obtained through a stochastic background search by PTAs (light gray) and the CMB (dark gray). The dashed line (light blue) is the accessible parameter space predicted for the SKA. The dotted line (dark blue) is the prediction for the SKA from the clustering of cosmic string loops in our Galaxy's halo. }
  \label{Fig:constraint3}
  \end{center}
\end{figure}

Finally, in Figure~\ref{Fig:constraint3}, we present the constraint in the $G\mu$ -- $\alpha$ plane, compared with constraints by other types of observations; cosmic microwave background (CMB) \citep{Ade2014} and a stochastic GW background\footnote{We derived the shaded area using the upper limit on the stochastic GW background for different spectral index given in \cite{Arzoumanian}; $A_{\rm GWB}=1.45\times 10^{-15}$ for $\gamma=13/3$ and it scales as $\propto 10^{-0.4 \gamma}$, where the characteristic strain amplitude is parametrized as $h_c(f)=A_{\rm GWB}\left(\frac{f}{\rm yr^{-1}}\right)^\zeta$ and $\gamma=3-2\zeta$.}.  Although the stochastic background provides better constraint on the string parameters, we stress that the burst GW search provides independent constraints.

\subsection{Prediction for the IPTA and future SKA data sets} \label{sec:prediction for IPTA}

As mentioned in section \ref{sec:results}, we obtain large values for the detection statistic at MJD 53250, 53500 and 54750. The events at MJD 53250 and 53500 are expected to be spurious since they do not appear in any PPTA pulsars other than PSR~J1939+2134. Whereas, even though the event with $D_{\rm max,sky} = 25.7$ for $W = 75$~days at MJD 54750 is likely to be spurious and caused by the effect of a receiver configuration change at that time, it is an event that is detected in the timing residuals of more than one pulsar.
If it is caused by the GW burst, the effect should appear in the timing residuals for pulsars in the Northern hemisphere. Here we provide a prediction for the IPTA pulsars. Table \ref{tab:prediction ipta} shows the expected effect of the GW burst, which yields $D_{\rm max,sky}=25.7$ with $W = 75$ days at MJD~54750, in IPTA pulsars that are not observed by the PPTA but are observed by the NANOGrav and EPTA. The second column in the table represents the expected maximum post-fit timing residuals at MJD 54700. Note that the timing residual induced by the GW burst is maximized at not the epoch but a little before (or after) it as seen in Figure~\ref{fig:example_burst}. The third and forth columns are the weighted RMS of the IPTA residuals (obtained from \citealt{perera19}) and the signal to noise ratio defined as $|r(t_{\rm max})|/\sigma_{\rm{w}}$, respectively. We find that the strongest effect of the GW burst should appear in PSR~\textcolor{red}{J1640$+$2224}. In order to verify whether $D_{\rm max,sky}=25.7$ is a GW signal or not, IPTA data, especially PSR~\textcolor{red}{J1640$+$2224}, will be valuable to study.

\begin{table}
	\centering
	\caption{Expected sensitivities of IPTA pulsars for the GW burst causing $D=25.7$. {\it Second column}: Expected post-fit timing residuals at $t_{\rm max}$ = MJD 54750 where variation in the residuals is the largest. {\it Third column}: Weighted RMS of the residuals \citep{perera19}. {\it Forth column}: Signal to noise ratio defined as $|r(t_{\rm max})|/\sigma_{\rm{w}}$.}
	\begin{tabular}{cccc}
		\hline
		Pulsar Name & $r(t_{\rm max})$ [$\mu$s] & $\sigma_{\rm{w}}$ [$\mu$s] & S/N \\
		\hline
		J0030$+$0451 & $-$5.4$\times 10^{-2}$ & 1.48 & 3.7$\times 10^{-2}$ \\
		J0034$-$0534 & 2.0$\times 10^{-3}$ & 4.19 & 4.8$\times 10^{-4}$ \\
		J0218$+$4232 & 1.1 & 7.01 & 0.16 \\
		J0610$-$2100 & 0.13 & 4.88 & 2.6$\times 10^{-2}$ \\
		J0621$+$1002 & 0.17 & 6.57 & 2.6$\times 10^{-2}$ \\
		J0751$+$1807 & $-$0.81 & 3.00 & 0.27 \\
		J0900$-$3144 & $-$0.49 & 3.21 & 0.15 \\
		J1012$+$5307 & $-$1.6 & 1.91 & 0.86 \\
		J1455$-$3330 & 0.39 & 4.12 & 0.10 \\
		J1640$+$2224 & 1.1 & 0.77 & 1.4 \\
		J1721$-$2457 & 0.34 & 12.21 & 2.8$\times 10^{-2}$ \\
		J1738$+$0333 & 0.47 & 1.38 & 0.34 \\
		J1751$-$2857 & 0.20 & 2.85 & 6.9$\times 10^{-2}$ \\
		J1801$-$1417 & 0.22 & 2.76 & 7.9$\times 10^{-2}$ \\
		J1802$-$2124 & 0.18 & 2.76 & 6.7$\times 10^{-2}$ \\
		J1804$-$2717 & 0.15 & 3.72 & 4.1$\times 10^{-2}$ \\
		J1843$-$1113 & $-$3.7$\times 10^{-2}$ & 0.71 & 5.3$\times 10^{-2}$ \\
		J1853$+$1303 & $-$0.19 & 1.31 & 0.14 \\
		J1910$+$1256 & $-$0.35 & 1.42 & 0.25 \\
		J1911$+$1347 & $-$0.36 & 4.30 & 8.5$\times 10^{-2}$ \\
		J1911$-$1114 & $-$0.21 & 1.09 & 0.20 \\
		J1918$-$0642 & $-$0.29 & 1.80 & 0.16 \\
		J1955$+$2908 & $-$0.91 & 3.20 & 0.28 \\
		J2010$-$1323 & $-$0.50 & 2.53 & 0.20 \\
		J2019$+$2425 & $-$1.0 & 9.64 & 0.11 \\
		J2033$+$1734 & $-$1.0 & 13.65 & 7.6$\times 10^{-2}$ \\
		J2229$+$2643 & $-$1.2 & 4.28 & 0.29 \\
		J2317$+$1439 & $-$0.75 & 0.87 & 0.87 \\
		J2322$+$2057 & $-$0.79 & 6.74 & 0.12 \\
		\hline
	\end{tabular}
	\label{tab:prediction ipta}
\end{table}

The Square Kilometre Array (SKA) telescope will observe a large number of millisecond pulsars, with high cadence and high timing precision.  In Figure~\ref{Fig:constraint3}, we also show the prediction assuming that the SKA can reach $h^{\rm CS,peak}=10^{-16}$. Such data sets would improve the constraints up to $G\mu \sim 10^{-10}$ (dashed light-blue line) and, if string loops are clustered in the dark-matter halo of our Galaxy \citep{Chernoff:2017fll}, we may even be able to reach down to $G\mu \sim 10^{-12}$ for large $\alpha$ (dotted dark-blue line). Usually, stochastic background searches provide better constraints on the string parameters at PTA frequencies. However, if the string tension is low, $10^{-15}<G\mu<10^{-8}$, the loop lives longer and the old loops tend to cluster, enhancing the local number density of loops up to $\sim 10^5$. Thus, we can access the parameter space of small $G\mu$ with the improved sensitivity of the SKA, which may enhance the possibility of single burst detection. A tight bound on the string tension by the SKA would be extremely useful to test models of cosmic superstrings such as the KKLMMT  \citep{kachru03} scenario, where the tension is predicted to be in the range $10^{-12}<G\mu<10^{-6}$ \citep{Jones:2002cv,Sarangi,Jones}.

\section{Conclusion} \label{sec:conclusion}

For this work, we developed an algorithm for detection of a GW burst from a cosmic string as a single source event and applied that algorithm to the PPTA data set. 
We found four detections with a false alarm probability less than 1\%. Three of these appear as ``detections'' of the pulsar term in PSR~J1939+2134 at a time when there is a large data gap in the observations of that pulsar. We attribute these apparent detections to incorrect red noise modelling because of the data gap, and we remove that pulsar from the analysis of epochs centered on MJDs 53250 and 53500. The fourth apparent detection, at MJD 54750, involves two pulsars PSRs~J1939+2134 and J0437$-$4715. We believe that this detection is also spurious because it is coincident with a receiver re-configuration, but it remains interesting. We removed both pulsars from analysis of the epoch centered on MJD~54750 and reanalyzed the data set. This removed all the apparent detections and we use this edited data set to place constraints on the cosmic string tension as a function of the burst width, which corresponds to the loop size. We found that, for two reasons, the constraint on the string tension becomes more stringent as the width increases. One is that our data sets are more sensitive for GW amplitudes from wider events, and another is that larger loops emit stronger GW bursts. The constraint on the tension turns out to be weaker than the other types of probes, such as the CMB and the stochastic background search by the PTA. However, we stress that this is an independent test for the existence of cosmic string GWs at the nano-Hertz frequency.

In the future, the SKA will increase the sensitivity of the PTA and will allow us to access the parameter space of small $G\mu$ up to $\sim 10^{-10}$. Furthermore, such data sets will provide an independent means to test for the clustering of cosmic string loops in our Galaxy, which cannot be tested either through the CMB analysis or from a stochastic background search.

\section*{Acknowledgements}

The Parkes radio telescope is part of the Australia Telescope National Facility which is funded by the Commonwealth of Australia for operation as a National Facility managed by CSIRO. This paper includes archived data obtained through the Parkes Pulsar Data archive on the CSIRO Data Access Portal (http://data.csiro.au). NY was financially supported by JSPS KAKENHI Grant Numbers JP19J11575  and the Grant-in-Aid from the Overseas Challenge Program for Young Researchers of JSPS. SK is partially supported by JSPS KAKENHI No.17K14282 and Career Development Project for Researchers of Allied Universities. KT is partially supported by JSPS KAKENHI Grant Numbers JP15H05896, JP16H05999 and JP17H01110, and Bilateral Joint Research Projects of JSPS. NT is a Jansky fellow of the National Radio Astronomy Observatory.  XZ, MB, PL, JW, DR, RS, LW are supported by ARC CE170100004. JW is supported by the Youth Innovation Promotion Association of Chinese Academy of Sciences, the National Key R\&D Program of China (No. 2017YFA0402602), and the Strategic Priority Research Program (B) of the Chinese Academy of Sciences (No. XDB230102000).   RS acknowledges support from the Australian Research Council grant FL150100148.

\section{Data availability}

The observations from the Parkes radio telescope are publicly available from \url{https://data.csiro.au/} after an 18 month embargo period. Note that all data included in this paper are out of this embargo period and are available using the P456 observing project code.  These data were processed and those data products (which form the basis of this paper) are available from \url{https://doi.org/10.25919/5bc67e4b7ddf2}.






\appendix
\section{Event rate of GW bursts from cosmic string cusps}
Here we outline the calculation of event rate, which follows the method in \cite{Siemens} (see also \citep{Kuroyanagi:2012,Kuroyanagi:2013}). For string network evolution, we use the velocity-dependent one-scale
model where all loops are assumed to be formed with the same size. 
Given that the cosmic string is extremely thin, with the diameter scale of the same order of magnitude as that of a proton and therefore much smaller than any cosmological scale, the cosmic strings are considered one-dimensional objects and studied in the zero-width, or Nambu-Goto approximation. This action is classically equivalent to Polyakov action for the bosonic superstring theory. These cosmic strings are often called Nambu-Goto strings.
In our model, the string network of infinite strings is characterized by
a correlation length $\xi$.  The total length $L$ of infinite strings
in volume $V$ is given by $L=V/\xi^2$, and the average string energy
density is given by $\rho=\mu/\xi^2$.  Defining
$\gamma\equiv\xi/t$, the equation for energy conservation gives
\begin{equation}
\frac{t}{\gamma}\frac{d\gamma}{dt}=-1+\nu +\frac{\tilde{c}pv}{2\gamma}+\nu v^2,
\label{gammaeq}
\end{equation}
while the equations of motion for the Nambu-Goto string yields an
equation for the evolution of the typical root-mean-square velocity
$v$ of infinite strings,
\begin{equation}
\frac{dv}{dt}= (1-v^2)H (\frac{k (v)}{\nu\gamma}-2v),
\label{veq}
\end{equation}
where $k(v)=\frac{2\sqrt{2}}{\pi}\frac{1-8v^6}{1+8v^6}$, $H\equiv
\dot{a}/a$ is the Hubble expansion rate, and the scale factor $a$ is
parametrized as $a (t)\propto t^{\nu}$.  The third term in the
right-hand side of Eq.\eqref{gammaeq} represents the loss of energy
from infinite strings by the production of loops.  The constant
parameter $\tilde{c}$ represents the efficiency of loop formation and
is set to be $\tilde{c}=0.23$.  The reconnection probability $p$,
which could be much less than $1$ for cosmic superstrings, decreases
the loop formation rate and reduces the value of $\gamma$ for smaller
$p$. The above sets of equations have asymptotic solutions. For $p=1$, we obtain $\gamma_r=0.27$ for the radiation-dominated era and $\gamma_m=0.62$ for the matter-dominated era. 

The string loops are formed by the reconnection of long strings with 
themselves, that oscillate relativistically and eventually lose their energy to GWs. We parametrize the initial loop size as $\alpha=l_b/t_b$, where $b$ denotes the time of loop formation. After their birth, loops shrink by GW emission and the loop size
at time $t$ is given by
\begin{equation}
  l(t) = \alpha t_b - \Gamma G\mu (t-t_b),
  \label{eq:l}
\end{equation}
where $\Gamma$ is a numerical constant which represents the efficiency of GW emission and we take $\Gamma=50$.  The number density of loops created at  time $t_b$ is given by
\begin{equation}
  \frac{dn}{dt_b} = \frac{C}{\alpha p^2 t_b^4} \,,
  \label{eq:dndtb}
\end{equation}
where the coefficient $C$ is typically determined by the number of
infinite strings in the string network. In this paper, we take $C=1/\gamma^2$, which gives $C_r=1/\gamma_r^2=13.7$ for the radiation-dominated era and $C_m=1/\gamma_m^2=2.63$ for the matter-dominated era. From Eq. \eqref{eq:l}, we obtain
\begin{equation}
  t_b = \frac{l + \Gamma G\mu t}{\alpha + \Gamma G\mu},
  \label{eq:tb}
\end{equation}
and
\begin{equation}
  \frac{dt_b}{dl} = \frac{1}{\alpha + \Gamma G\mu}.
  \label{eq:dtbdl}
\end{equation}
Using Eqs. \eqref{eq:dndtb} and \eqref{eq:dtbdl} and taking into account the fact that the number density decreases as $\propto a^{-3}$ by the expansion of the Universe, we find the number density of loops with length between $l$ and $l+{\rm d} l$ at time $t$ is given as
\begin{equation}
  \frac{dn}{dl}(t) = \frac{C}{\alpha p^2} \frac{1}{(\alpha + \Gamma G\mu)t_b^4}
  \left(\frac{a(t_b)}{a(t)}\right)^3  \,.
  \label{eq:number}
\end{equation}

The rate of GWs coming towards us is obtained by multiplying the number of cusp events per time $2N_c/[(1+z)l]$. We assume that cusps happen once every oscillation period, namely $N_c=1$. We also take into account the beaming effect $\frac{1}{4}\theta_m^2$, where $\theta_m\equiv[(1+z)f \cdot l]^{-1/3}$ is the maximum angle between the line of sight and the direction of a cusp velocity to detect a GW of frequency $f$. Multiplying by the volume $\frac{dV}{dz}=\frac{4\pi a(z)^2 r(z)^2}{H(z)(1+z)}$, the number of GW events between $z$ and $z+dz$ is given by
\begin{equation}
  \frac{dR}{dzdl}
  = \frac{1}{4}\theta_m(f,z,l)^2\frac{2N_c}{(1+z)l}\frac{dn}{dl}\frac{dV}{dz}
  \Theta(f \cdot l(1+z)-2) \,,
  \label{eq:dRdzdl}
\end{equation}
where the Heaviside step function represents a low-frequency cut-off below $f_l=2/l$ at the emission. From Eq. \eqref{eq:amp}, we find
\begin{equation}
  l = \left( \frac{\tilde{h}(f)r(z)}{G\mu}(1+z)^{1/3}f^{4/3}\right)^{3/2} \,,
  \label{eq:l2}
\end{equation}
and
\begin{equation}
  \frac{dl}{d\tilde{h}} = \frac{3l}{2\tilde{h}} \,.
  \label{eq:dldh}
\end{equation}
Using Eqs. \eqref{eq:number} and \eqref{eq:dldh}, we can rewrite Eq. \eqref{eq:dRdzdl} as
\begin{eqnarray}
\frac{dR}{dzd\tilde{h}} 
& = & \frac{3}{4}\theta_m(f,z,l)^2\frac{N_c}{(1+z)\tilde{h}}\frac{C}{\alpha p^2} \frac{1}{(\alpha + \Gamma G\mu)t_b^4}
\nonumber\\
& ~ & \times\left(\frac{a(t_b)}{a(t)}\right)^3 \frac{dV}{dz}\Theta(f \cdot l(1+z)-2) \,.
\label{eq:dRdzdh2}
\end{eqnarray}








\bsp	
\label{lastpage}

\end{document}